\documentclass[a4paper ]{article}
\usepackage[english]{babel}
\usepackage{graphicx}
\usepackage{amsmath}
\usepackage{amssymb}
\usepackage{bm}

\begin{document}

\title{Zeroth-order considerations on an accelerator-based gravitational wave amplifier}

\author{Velizar Miltchev\thanks{velizar.miltchev@desy.de}}

\maketitle

\begin{abstract}
The operation of an accelerator-based gravitational wave amplifier has been studied, taking into account the interaction between beam and gravitational wave and neglecting all effects related to the emission of - and interaction with electromagnetic radiation. The small-gain operation mode has been considered. The gravitational counterpart of the Madey theorem as well some basic amplifier quantities as resonance wavelength, amplifier gain and gain length have been  derived. 
\end{abstract}

\section{Introduction}
\begin{figure}[h!]   
\begin{center}
{ \includegraphics[angle=0,width=1\columnwidth]{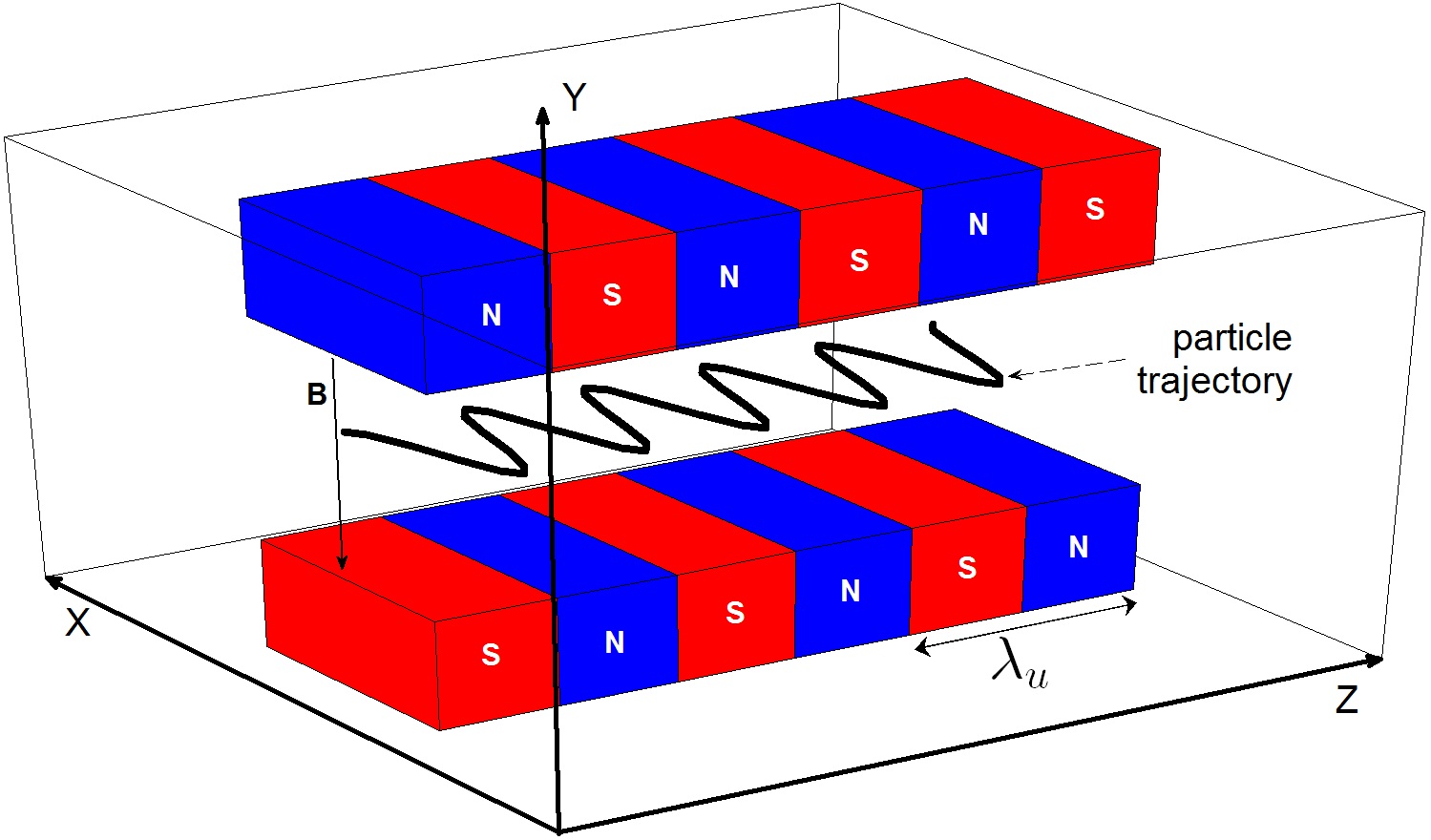}}
\caption{Sketch of  the gravitational wave amplifier set-up.}
\label{fig:UndulatorSetUp}
\end{center}
\end{figure}

The set-up of the considered gravitational wave amplifier is identical to a free-electron laser (FEL) as shown in Fig.\ref{fig:UndulatorSetUp}. It consists of an arrangement of permanent magnetic dipoles called undulator and a relativistic charged particle beam, which propagates along the axis ($z$-axis) of the undulator.
The polarity of undulator dipoles alternates  along the z-axis with a period $\mathrm{\lambda_u}$ so that they create a static magnetic filed, which ideally is given by $\mathrm{\textbf{ B}=\left(0, B\sin\left(k_uz \right), 0 \right)}$. Here $\mathrm{k_u}=2\pi/\lambda_u$  and the non-zero component of the magnetic field points along the $y$-axis. For to facilitate the following discussion it is helpful to write explicitly the electromagnetic tensor for the undulator field:
\begin{gather}
 \begin{bmatrix}{ F^\mu}_\nu 
\end{bmatrix}
 = 
  \begin{bmatrix}
   0& 0& 0& 0\\
   0& 0& 0&  - B\sin\left(k_uz \right) \\
   0& 0& 0&  0 \\
   0&  B\sin\left(k_uz \right)& 0&  0
   \end{bmatrix}
\label{eq:FieldUndNoGW}
\end{gather}
The equation of motion in the rest frame of the undulator for a beam particle of charge $q$, Lorentz factor $\gamma$  and rest mass $m$ is:
\begin{gather}
\gamma m \frac{d}{dt}\bm{u}^\mu=q{ F^\mu}_\nu\bm{u}^\nu
\label{eq:EqMotionNoGW}
\end{gather}
here $\mu, \nu = 0, 1, 2, 3$ and $[\bm{u}^\mu]=\gamma\left(c, v_x, v_y, v_z\right)$ is the 4-velocity of the particle. A solution of (\ref{eq:EqMotionNoGW}) for $\mu = 1, 2, 3$ can be found elsewhere \cite{UVFELs}:
\begin{gather}
\bm{u}^1 =cK\cos\left( k_uz\right)
\label{eq:VX}
\end{gather}
\begin{gather}
\bm{u}^3 = \gamma\bar \beta c - \frac{cK^2}{4\gamma}\cos\left( 2\bar \beta ck_u t\right)
\label{eq:VZ}
\end{gather}
where the following abbreviations have been introduced:
\begin{gather}
K =\frac{qB}{mck_u}
\label{eq:Kparam}
\end{gather}

\begin{gather}
\bar \beta = 1-\frac{1}{2\gamma^2}\left(1+\frac{K^2}{2} \right)
\label{eq:BetaBar}
\end{gather}
The particle motion along the $y$-axis is unaffected by the undulator field. Therefore, for simplicity, the vertical velocity $\bm{u}^2$ and position will be assumed zero.  For $\mu = 0$ equation (\ref{eq:EqMotionNoGW}) yields $\gamma = \mathrm{const}$,

\begin{gather}
mc\gamma\frac{d\gamma}{dt}=0
\label{eq:DgamNoEField}
\end{gather}
exactly as expected in the absence of electric field.\\
Essential for the amplifier operation is that massive particles undergo transverse motion. Therefore in the following, in order to be able to distinguish the effects related to the gravitational wave the particles will be assumed with no electric charge and thus all effects related to the emission of - and interaction with electromagnetic radiation as well as space charge effects are going to be neglected.

\section{Energy exchange between particles and gravitational wave}
In the following, a plane linearly polarized gravitational wave traveling along the $z$-axis in the same direction as the particle beam will be considered. The gravitational wave of wavelength $\lambda$ has the form:
\begin{gather}
h_{\mu \nu} = A_{\mu \nu}\cos\left( kz -\omega t\right)
\label{eq:GW}
\end{gather}
where $k=2\pi/\lambda$ and the amplitude tensor is:
\begin{gather}
 \left[A_{\mu \nu}\right]= \mathrm{diag}\left(0, -a, a, 0 \right)
\label{eq:AmplTens}
\end{gather}
with $\lvert a\rvert\ \ll 1 $. Even though $a$ may vary along the undulator, the rate of change is very small and therefore it will be considered constant in the first order.  The metric tensor then becomes:
\begin{gather}
g_{\mu \nu} = \eta_{\mu \nu} + h_{\mu \nu}
\label{eq:MetricG}
\end{gather}
 where $\left[ \eta_{\mu \nu}\right] = \mathrm{diag}\left( 1, -1, -1, -1\right)$. In the new metric, equation (\ref{eq:DgamNoEField}) changes to:
\begin{gather}
mc\gamma\frac{d\gamma}{dt} + m{\Gamma^0}_{\mu\nu}\bm{u}^\mu \bm{u}^\nu=0
\label{eq:DgamGA}
\end{gather}
The metric connection is given by:
\begin{gather}
{\Gamma^\sigma}_{\mu\nu}=\frac{1}{2} \eta^{\sigma \rho}\left( \partial_\nu h_{\rho\mu} + \partial_\mu h_{\rho\nu}-\partial_\rho h_{\mu\nu}\right)
\label{eq:MertConnGen}
\end{gather}
Considering (\ref{eq:AmplTens}), in the case for $\sigma = 0$ there are only two nonzero connection coefficients:
\begin{gather}
{\Gamma^0}_{11}=-\frac{1}{2}\partial_0 h_{11}=-\frac{1}{2}\frac{\partial}{\partial\left( ct\right)}\left(-a\cos\left( kz -\omega t\right) \right)=\frac{a\omega}{2c}\sin\left(  kz -\omega t \right)
\label{eq:Gamm011}
\end{gather}
and
\begin{gather}
{\Gamma^0}_{22}=-\frac{1}{2}\partial_0 h_{22}=-\frac{1}{2}\frac{\partial}{\partial\left( ct\right)}\left(a\cos\left( kz -\omega t\right) \right)=-\frac{a\omega}{2c}\sin\left( kz -\omega t \right)
\label{eq:Gamm022}
\end{gather}
The gravitational wave has very small effect on the particle motion. Therefore it is safe to assume that equations (\ref{eq:VX}) and (\ref{eq:VZ}) will excellently approximate the velocities when the particles co-propagate with the gravitational wave. Taking into account that $\bm{u}^2=0$ and plugging (\ref{eq:VX}) and (\ref{eq:Gamm011}) into (\ref{eq:DgamGA}) yields:
\begin{gather}
\frac{d\gamma}{dt} = -\frac{a\omega K^2}{2\gamma}\sin\left( kz -\omega t\right)\cos^2\left( k_uz\right)
\label{eq:DgamFinal}
\end{gather}
Equation (\ref{eq:DgamFinal}) governs the energy exchange between a particle and the gravitational wave.

\section{Resonance condition}
A sustained energy transfer from or to the gravitational wave can happen only when the right-hand side of (\ref{eq:DgamFinal}) on average does not vanish along the undulator. The necessary conditions can be obtained by recasting the term in a different form:
\begin{gather}
\frac{a\omega K^2}{2\gamma}\sin\left( kz -\omega t \right)\cos^2\left( k_uz\right)=\frac{a\omega K^2}{8\gamma}\left [\sin \left( \Psi \right) +  \sin \left( \Xi \right) + 2 \sin \left( \alpha \right)\right ]
\label{eq:PSIintro}
\end{gather}
where the following phases have been defined:
\begin{gather}
\Psi = \left( k + 2k_u\right)z- \omega t 
\label{eq:PSIdef}
\end{gather}
\begin{gather}
\Xi = \left( k - 2k_u\right)z-\omega t=\Psi - 4k_uz
\label{eq:XIdef}
\end{gather}
\begin{gather}
\alpha = kz- \omega t =\Psi - 2k_uz
\label{eq:Alphadef}
\end{gather}
The phase $\Psi$ is the analogue of the ponderomotive phase known from the FEL-amplifiers theory \cite{UVFELs}. The condition $\Psi=\mathrm{const}$ implies:
\begin{gather}
\frac{d}{dt}\Psi = \left( k + 2k_u\right)\frac{d}{dt}z- \omega =\left( \frac{2\pi}{\lambda} + 2\frac{2\pi}{\lambda_u}\right)\bar \beta c-\frac{2\pi}{\lambda}c =0
\label{eq:PsiConstCond}
\end{gather}
In the last line the fast-oscillating term in equation (\ref{eq:VZ}) has been neglected, assuming $z(t) = \bar \beta ct$. After plugging $\bar\beta$ from (\ref{eq:BetaBar}) and solving for $\lambda$ one obtains the required resonance condition for the gravitational wavelength:
\begin{gather}
\lambda=\frac{\lambda_u}{4\gamma^2}\left( 1+\frac{K^2}{2}\right)
\label{eq:ResCond}
\end{gather}
where an ultra-relativistic ($\gamma \gg K$) particle beam has been assumed.\\
If $\Psi=\mathrm{const}$, the terms with $\Xi$ and $\alpha$ in (\ref{eq:PSIintro}) oscillate four and two times per undulator period respectively and on average do not contribute to the overall energy transfer. One can also show that only the phase $\Psi$ can remain constant i.e. the equations $d\Xi/dt=0$ and $d\alpha/dt=0$ do not have physically meaningful solution.\\
The resonance energy $\gamma_r$  is the particle energy at which equation  (\ref{eq:ResCond}) exactly holds for a given $\lambda$:
\begin{gather}
\gamma_r=\sqrt{\frac{\lambda_u}{4\lambda}\left( 1+\frac{K^2}{2}\right)}
\label{eq:GammaR}
\end{gather}
Finally, it is convenient to rewrite  (\ref{eq:DgamFinal}) in terms of $\Psi$ and $z$:
\begin{gather}
\frac{d\gamma}{dz} = -\frac{a\omega K^2}{8\gamma c}\sin\left(\Psi \right)
\label{eq:DgamPsiZ}
\end{gather}
 
\section{Longitudinal phase-space oscillations}
As discussed above, when $\gamma=\gamma_r$ the resonance condition is exactly fulfilled, $\Psi$ is constant and hence:
\begin{gather}
\frac{d\Psi}{dz}(\gamma_r)=\left( k + 2k_u\right) - \frac{\omega}{\bar \beta(\gamma_r) c}= 0
\label{eq:dPsiDzGamR}
\end{gather}
In the last line $\bar \beta$ has been explicitly written as a function of $\gamma$ and the substitution $t = z/(\bar \beta c)$ has been made. If a particle has a slightly different energy $\gamma = \gamma_r +\Delta \gamma$, then to first order in $\Delta \gamma$ the rate of change of $\Psi$ is given by:
\begin{gather}
\begin{split}
\frac{d\Psi}{dz}(\gamma_r +\Delta \gamma)&\approx\frac{d\Psi}{dz}(\gamma_r) +  \left. \Delta \gamma\frac{d}{d\gamma}\left( \frac{d\Psi}{dz}\right) \right |_{\gamma=\gamma_r}\\
&=  \left. \Delta \gamma\left(\frac{\omega}{{\bar\beta}^2c}\frac{d\bar\beta}{d\gamma}\right) \right |_{\gamma=\gamma_r}
\label{eq:dPsiDzDeltaGam}
\end{split}
\end{gather}
which on using (\ref{eq:BetaBar}), (\ref{eq:ResCond}) and assuming again $\bar \beta \approx 1$  transforms to
\begin{gather}
\frac{d\Psi}{dz}(\gamma_r +\Delta \gamma)=4k_u\eta
\label{eq:dPsiDzFinal}
\end{gather}
where the relative energy deviation has been defined as $\eta =\left(\gamma - \gamma_r\right)/{\gamma_r}$. Equation (\ref{eq:dPsiDzFinal}) together with (\ref{eq:DgamPsiZ}) forms a system of coupled differential equations.
\begin{gather}
\begin{split}
\frac{d\eta}{dz} &= -\frac{a\omega K^2}{8\gamma_r^2 c}\sin\left(\Psi \right) \\
\frac{d\Psi}{dz}&=4k_u\eta
\end{split}
\label{eq:SystemPendulum}
\end{gather}
On combining them one obtains a second order differential equation, which is formally identical to the pendulum equation occurring in the low-gain FEL theory \cite{UVFELs, Wiedemann}: 
\begin{gather}
\frac{d^2\Psi}{dz^2}+\Omega_g^2\sin(\Psi)=0~~\mathrm{with} ~~\Omega_g^2=\frac{a\omega K^2k_u}{2c\gamma_r^2}
\label{eq:Pendulum}
\end{gather}

\section{Gravitational analogue of Maday theorem}
The conventional electromagnetic FEL-amplifiers operating in the low-gain \\regime are typically installed at storage rings inside an optical cavity. The electromagnetic wave remains captured in the cavity and gets amplified over many revolutions. Thus the output field amplitude exceeds the initial one by many orders of magnitude even though the gain per passage might be a few percents only. If, as for the gravitational waves, suitable mirrors are not available, then one can think of high-gain operation mode, in which the output amplitude increases exponentially with $z$ and just one passage is required.
Nevertheless, in the following a gravitational wave cavity-amplifier operating in the low-gain regime is considered. The amplifier gain $\Gamma_g$ is defined as the relative energy transfer from the beam particles to the gravitational wave over a single pass through the undulator:
\begin{gather}
\Gamma_g=-\frac{Nmc^2\Delta\gamma}{V\rho_g}=-\frac{nmc^2\Delta\gamma}{\rho_g}~~\mathrm{with} ~~\rho_g=\frac{c^2a^2\omega^2}{32\pi G}
\label{eq:GainFirst}
\end{gather}
with $N$ - the total number of beam particles, $n$ - particle density, $V$ - volume occupied by the gravitational wave, $\rho_g$ - energy density of the gravitational wave and $G$ - gravitational constant. For simplicity it is assumed that $V$ coincides with the volume occupied by the beam. Following the arguments in \cite{Wiedemann}, the total energy loss  $\Delta \gamma$  for one particle can be calculated from  (\ref{eq:SystemPendulum}) and (\ref{eq:Pendulum}) applying second order perturbation theory,
\begin{gather}
\Delta \gamma=\frac{\gamma_rL^3\Omega_g^4}{32k_u}\frac{d}{d\xi}\left( \frac{\sin\xi}{\xi}\right)^2
\label{eq:DeltaGammaWied}
\end{gather}
where $L$ is the undulator length and $\xi=2k_uL\eta$. 
Substituting the relations (\ref{eq:DeltaGammaWied}) and (\ref{eq:Kparam}) in (\ref{eq:GainFirst}) one obtains\footnote{for simplicity's sake, higher-order effects related to variation of longitudinal velocity are neglected here.}
\begin{gather}
\Gamma_g=-\frac{G n q^4 L^3 B^4 \lambda_u^3}{32 \pi^2 m^3 c^6\gamma_r^3}\frac{d}{d\xi}\left( \frac{\sin\xi}{\xi}\right)^2
\label{eq:GainFinal}
\end{gather}
It is convenient to introduce the the power gain length $L_g$
\begin{gather}
L_g=\frac{1}{\sqrt{3}}\left(     \frac{8 c^6 m^3 \pi^2 \gamma_r^3}{B^4 Gnq^4\lambda_u^3}      \right)^{1/3}
\label{eq:LG}
\end{gather}
Thus (\ref{eq:GainFinal}) can be recast in the form
\begin{gather}
\Gamma_g=-\frac{1}{12\sqrt{3}}\left(\frac{L}{L_g}\right)^3\frac{d}{d\xi}\left( \frac{\sin\xi}{\xi}\right)^2
\label{eq:GainWithLG}
\end{gather}
Equation (\ref{eq:GainFinal}) is the gravitational counterpart of the Madey theorem, which was first derived for electromagnetic wave amplifiers \cite{Madey}. The corresponding energy gain for an  electromagnetic wave is given by:
\begin{gather}
\Gamma_{em}=-\frac{n q^4 B^2 L^3 \lambda_u}{16\pi c^4 \epsilon_0m^3 \gamma_r^3}\frac{d}{d\xi}\left( \frac{\sin\xi}{\xi}\right)^2
\label{eq:GainEM}
\end{gather}
and hence
\begin{gather}
\frac{\Gamma_g}{\Gamma_{em}} =\frac{G\epsilon_0}{2\pi c^2}\left( B\lambda_u\right)^2
\label{eq:GainRatio}
\end{gather}
The gain ratio in (\ref{eq:GainRatio}) is independent on the particles type and depends only on physical constants and undulator properties. This gives a possibility to roughly estimate the gain length $L_g$ of the gravitational wave amplifier using the FEL gain length $L_{em}$:
\begin{gather}
L_g \propto L_{em}\sqrt[\leftroot{-1}\uproot{2}\scriptstyle 3]{ \frac{\Gamma_{em}}{\Gamma_g}}= L_{em}\sqrt[\leftroot{-1}\uproot{2}\scriptstyle 3]{ \frac{2\pi c^2}{G\epsilon_0 B^2 \lambda_u^2}}
\label{eq:SinglePassLG}
\end{gather}
Equation (\ref{eq:SinglePassLG}) bears interpretation. On the one hand, $ L_{em}\propto m$, so one expects lighter particles to produce shorter gain lengths, what makes $e^-$-accelerators more feasible. On the other hand, for to minimize EM-radiation effects, one needs heavy particles such as $p^+$ or even heavy ions. \\ As a numerical example one can take a peak field $B$ = 2\,T, undulator period $\lambda_u$ = 0.2\,m and FEL-gain length $ L_{em}$ = 1\,m. In this case  Eq.(\ref{eq:SinglePassLG}) yields $L_g\sim$ 10\,light-hours. \\Obviously such effective amplifier length can only be realized in a storage-ring, if at all. For example, similar to conventional ring lasers, one can imagine a storage-ring amplifier, but with the classical mirrors replaced by gain guiding.

\section{Summary}
The operation of an FEL-like gravitational wave amplifier has been studied in the framework of a simplified model, in which the effects related to the emission of - and interaction with electromagnetic radiation have been ignored. This made possible to derive the gravitational counterparts of some basic FEL-quantities like ponderomotive phase, resonance wavelength, amplifier gain and gain length.

\end{document}